\documentstyle[11pt,appb,psfig]{article}        %
\begin{document}
\title{Meson Production and Meson Properties at Finite Nuclear
Density\footnote{Supported by BMBF, FZ J\"ulich and GSI Darmstadt}}
\author{W. Cassing\thanks{In collaboration with E. L. Bratkovskaya, J. Geiss,
C. Greiner, U. Mosel and A. Sibirtsev} \\
Institut f\"ur Theoretische Physik, Universit\"at Giessen \\
D-35392 Giessen, Germany}
\date{ }
\maketitle

\begin{abstract}
The properties of $\pi, \eta, K^+$ and $K^-$ mesons are studied in
nuclear reactions from SIS to SPS energies within the covariant
transport approach HSD in comparison to the experimental data.  Whereas
the pion, $\eta$ and kaon abundancies and spectra indicate little or
vanishing selfe-nergies for these mesons in the medium, antikaons (as
well as antiprotons) are found to experience strong attractive
potentials in nucleus-nucleus collisions at SIS energies. However, even
when including these potentials the $K^+$ and $K^-$ spectra
at AGS energies are noticeably underestimated showing an experimental
excess of strangeness that points towards a nonhadronic phase in these
reactions.  On the other hand the $K^+, K^-$ production at SPS energies
is well described by the hadronic approach without incorporating any
parton degrees of freedom.
\end{abstract}

\section{Introduction}
The aim of high energy heavy-ion collisions at the GSI Schwerionen
Synchrotron (SIS), the Brookhaven Alternating Gradient Synchrotron
(AGS) and the CERN Super Proton Sychrotron (SPS) is to investigate
nuclear matter under extreme conditions, i.e. high temperature and
density. The most exciting prospect is the possible observation of a
signal for a phase transition from normal nuclear matter to a
nonhadronic phase, where partons are the basic degrees of freedom. In
this context strangeness enhancement in heavy-ion collisions compared
to proton-proton collisions has been suggested as a possible signature
for the phase transition \cite{Rafelski1}. On the other hand, precursor
effects might already be seen at SIS energies since densities up to
3$\times \rho_0$ can be achieved in central collisions of heavy nuclei
\cite{Cass} and the effect of meson potentials can be studied with a
higher sensitivity to the productions thresholds, respectively.
Furthermore, dilepton spectroscopy should be well suited to investigate
the in-medium properties of especially the $\rho$-meson which due to
its short life time preferentially decays in the medium \cite{CBPR98}.

In this contribution a brief survey is presented on the information
gained so far in comparison of experimental data to nonequilibrium
transport theory, here the Hadron-String-Dynamics (HSD) approach
\cite{Ehehalt}. For a more detailed discussion of the issues presented
the reader is refered to a recent review \cite{CBPR98}.

\section{Analysis of meson properties from SIS to SPS energies}
The production of particles especially at 'subthreshold' energies is
expected to provide valuable information about the properties of
hadrons at high baryon density and temperature \cite{Cass}. Their
relative abundance and spectra should reflect the in-medium properties
of the particles produced since for a 'dropping' mass -- i.e. a reduced
quasiparticle energy in the medium,
\begin{equation}
m^* = \omega({\bf p} = 0) = \sqrt{m_0^2 + \Pi_h(\rho_B, \rho_S,
{\bf p} = 0)},
\label{effm}
\end{equation}
where $\Pi_h(\rho_B, \rho_S, ..)$ denotes the meson self-energy as a
function of the baryon density $\rho_B$ and scalar density $\rho_S$ --
the particle can be created more abundantly. On the other hand it will
be suppressed in case of repulsive potentials. Furthermore, a
quasiparticle feeling an attractive potential at finite baryon density
will be decelerated during its propagation out of the medium and thus
asymptotically its momentum spectrum will be enhanced at low relative
momenta with respect to the baryon matter rest frame \cite{Brat98mt}.
The opposite holds in case of repulsive potentials.

\begin{figure}[t]
\phantom{a}\vspace*{-5mm}\hspace*{-0.5cm}
\psfig{figure=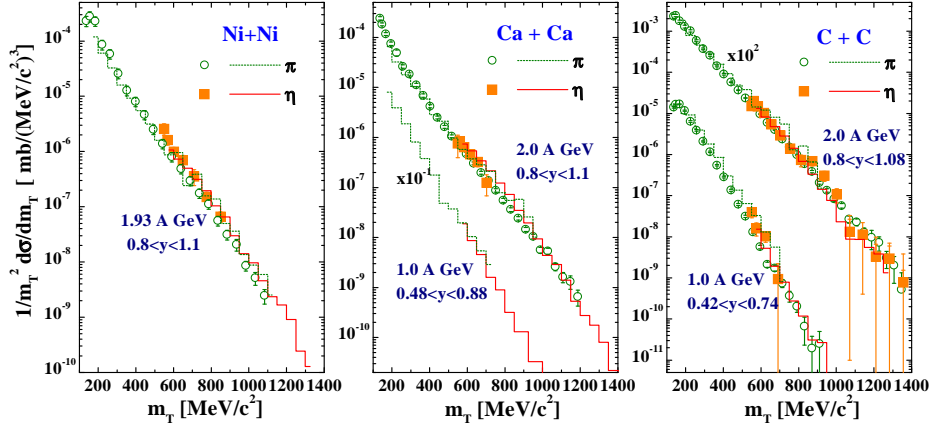,width=13cm}
\vspace*{-32mm}
\caption{The calculated transverse mass spectra for neutral pions and
$\eta$-mesons in comparison to the data from the TAPS Collaboration
(see text).}
\label{Fig1}
\end{figure}

As mentioned above the dynamics of hadron-hadron, hadron-nucleus and
nucleus-nucleus collisions is described within the HSD transport
approach \cite{Ehehalt} that so far has been well tested in a large
dynamical domain \cite{Cassing1}. We start with pion and $\eta$
dynamics at SIS energies.  The transverse-mass spectra of $\pi^0$ and
$\eta$ mesons in heavy-ion collisions up to 2 A$\cdot$GeV have been
measured by the TAPS Collaboration~\cite{TAPSold,TAPS-CC,TAPS-CaNi} and
a $m_T$ scaling has been observed for both mesons.  The same scaling
can be found in the HSD transport calculations when including no pion
and $\eta$-meson self-energies.  In Fig.~\ref{Fig1} we compare the
results of the calculation \cite{BCRW97} for the inclusive
transverse-mass spectra of $\pi^0$ and $\eta$ mesons  with the TAPS
data. The r.h.s. shows the $m_T$ spectra for $\pi^0$'s (dashed
histogram) and $\eta$'s (solid histogram) for C~+~C at 1.0~A$\cdot$GeV
in the rapidity interval $0.42 \le y \le 0.74$ and at 2.0~A$\cdot$GeV
for $0.8 \le y \le 1.08$.  The experimental data -- the open circles
and solid squares correspond to $\pi^0$ and $\eta$ mesons, respectively
-- are taken from Ref.~\cite{TAPS-CC}. The theoretical results as well
as the experimental data at 2.0 A$\cdot$GeV  here are multiplied by a
factor of $10^2$.  The middle part corresponds to Ca~+~Ca at 1.0
A$\cdot$GeV  for $0.48 \le y \le 0.88$ (multiplied by $10^{-1}$) and at
2.0 A$\cdot$GeV  for $0.8 \le y \le 1.1$ in comparison with the data
from Ref.~\cite{TAPS-CaNi}.  The l.h.s.  shows the calculated $m_T$
spectra for Ni~+~Ni at 1.93~A$\cdot$GeV for $0.8 \le y \le 1.1$ in
comparison with the data from Ref.~\cite{TAPS-CaNi}.  As seen from
Fig.~\ref{Fig1} the HSD transport model gives a reasonable description
of the $m_T$ spectra of pions and $\eta$'s as measured by the TAPS
Collaboration without incorporating any medium modifications for both
mesons. It is important to point out that these calculations are
parameter-free in the sense that all production cross sections for
$\eta$ mesons are extracted from experimental data in the vacuum and
the $\eta$-nucleon elastic and inelastic cross sections are obtained by
using detailed balance on the basis of an intermediate N(1535)
resonance. As shown in Ref. \cite{BCRW97} an attractive potential for
the $\eta$-meson would lead to a violation of the $m_T$ scaling found
experimentally. Without explicit representation we note that the pion
and $\eta$ spectra at AGS and SPS energies also do not indicate
sizeable in-medium effects \cite{CBPR98}.

Kaons and antikaons have shown to be more promising in this respect
\cite{Cassing2,Li97f}.  Since the real part of the actual $K^+$ and
$K^-$ self-energy $\Pi_h$ in (\ref{effm}) is quite a matter of debate we adopt
a more practical point of view and as a guide for the analysis use a
linear extrapolation of the form,
\begin{equation}
\label{kmass}
m^*_K(\rho_B) = m_K^0 \left(1 - \alpha \frac{\rho_B}{\rho_0}\right),
\end{equation}
with $\alpha_{\bar{K}} \approx $ 0.2-0.25 for antikaons and $\alpha_K
\approx -0.06$ for kaons.  Alternative fits to the antikaon
self-energies lead to different values for the parameter
$\alpha_{\bar{K}}$ in the range 0.1 $\leq \alpha_{\bar{K}} \leq 0.3$
(cf.  Ref. \cite{Schaffner}). The choice $\alpha_{\bar{K}} \approx $
0.2 leads to a fairly reasonable reproduction of the antikaon mass
from Refs.  \cite{Kaplan,Nelson,LiKo} and the results from Waas, Kaiser
and Weise \cite{waas}. In (\ref{kmass}) a momentum dependence
of the kaon or antikaon potential has been neglected for reasons of
numerical simplicity. The dispersion analysis of Sibirtsev et al.
\cite{Sibdr} shows that this is roughly fulfilled for the kaon
potential, however, the antikaon potential should be more strongly
momentum dependent.
\begin{figure}[t]
\phantom{a}\vspace*{-10mm}\hspace*{3mm}
\psfig{figure=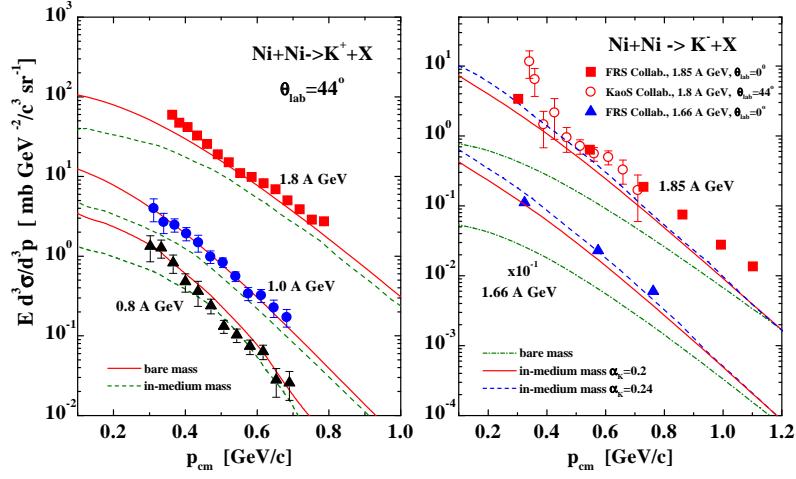,width=11cm}
\vspace*{-5mm}
\caption{ The calculated $K^+$ (l.h.s.) and $K^-$ (r.h.s.) momentum
spectra in the nucleus-nucleus cms for Ni + Ni reactions for different
meson potentials in comparison to the experimental data (see text).}
\label{Fig2}
\end{figure}

The Lorentz invariant $K^+$ spectra for Ni~+~Ni at 0.8, 1.0 and
1.8~A$\cdot$GeV are shown in Fig.~\ref{Fig2} (l.h.s.) in comparison to
the data from the KaoS Collaboration \cite{Senger}.  Here the full
lines reflect calculations including only bare $K^+$ masses ($\alpha_K
= 0$) while the dashed lines correspond to calculations with $\alpha_K
= - 0.06$ in Eq.~(\ref{kmass}), which leads to  an increase of the kaon
mass at $\rho_0$ by about 30~MeV.  The general tendency seen at all
bombarding energies is that the calculations with a bare kaon mass seem
to provide a better description of the experimental data for Ni~+~Ni
than those with an enhanced kaon mass. This trend continues to hold
also for the light system C + C as well as for the heavy systems Ru +
Ru and even Au + Au \cite{CBPR98}. Furthermore, this tendency is also
confirmed by the independent calculations from Li et al.  \cite{Li97f}.

On the other hand, the kaon flow in the reaction plane should show some
sensitivity to the kaon potential in the nuclear medium as put forward
by Li, Ko and Brown \cite{flow1,flow2}.  Here due to elastic scattering
with nucleons the kaons partly flow in the direction of the nucleons
thus showing a positive flow in case of no mean-field potentials
\cite{flow1}.  With increasing repulsive kaon potential the positive
flow will turn to zero and then become negative. In fact, experimental
data on kaon flow indicate a slightly repulsive potential for kaons in
the nuclear medium \cite{Brat622}. Further data with cuts on centrality
are expected to allow for more definite conclusions \cite{FOPI}.

We now turn to the production of antikaons which similar to antiprotons
\cite{Sib98} do clearly show the effect from attractive potentials in
the medium. We recall that for $\alpha_{\bar{K}}$ = 0 in Eq.
(\ref{kmass}) we recover the limit of vanishing antikaon self-energy,
whereas for $\alpha_{\bar{K}} \approx$ 0.2  we approximately describe
the scenario of Kaplan and Nelson~\cite{Kaplan,Nelson} or Waas, Kaiser
and Weise \cite{waas}. For practical purposes one should consider
$\alpha_{\bar{K}} $ to be a free parameter to be fixed in comparison to
the experimental data in order to learn about the magnitude of the
antikaon self-energy.  The $K^-$ spectra for Ni~+~Ni at 1.85 and 1.66
A$\cdot$GeV from Refs.~\cite{Schro,Gillitzer} are shown in Fig.
\ref{Fig2} (r.h.s.) for $\alpha_{\bar{K}}$ = 0, 0.2 and 0.24 where the
latter cases correspond to an attractive potential of $-100$ and
$-120$~MeV at density $\rho_0$, respectively.  We note, that due to the
uncertainties involved in the elementary $BB$ production cross sections
we cannot determine this value very reliably.  With increasing
$\alpha_{\bar{K}}$ not only the magnitude of the spectra is increased,
but also the slope becomes softer. This is most clearly seen at low
antikaon momenta because the net attraction leads to a squeezing of the
spectrum to low momenta. As seen from Fig. \ref{Fig2} the $K^-$ spectra
at 1.85 A$\cdot$GeV are underestimated at high antikaon momenta which
might indicate the necessity for explicit momentum dependent antikaon
potentials \cite{Sibdr}.

\begin{figure}[t]
\phantom{a}\vspace*{-10mm}
\psfig{figure=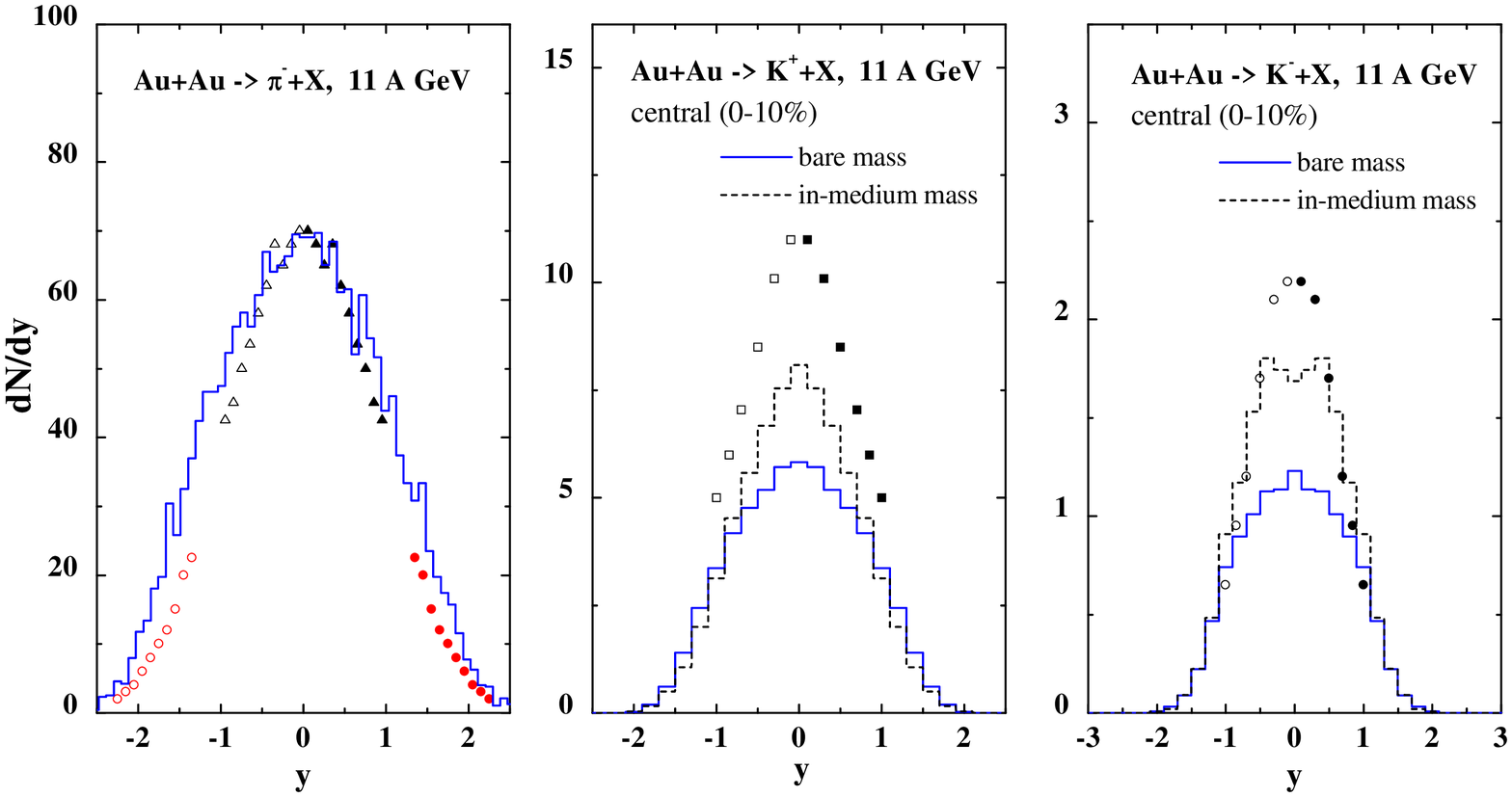,width=12.5cm}
\vspace*{-14mm}
\caption{The calculated $\pi^-$ (l.h.s.), $K^+$ (middle) and $K^-$ (r.h.s.)
rapidity spectra for central Au + Au reactions at 11 A$\cdot$GeV in
comparison to the experimental data. The solid histograms are
obtained without meson potentials whereas the dashed histograms represent
calculations with the same potentials as at SIS energies (see text). }
\label{Fig3}
\end{figure}

Whereas the kaon and antikaon dynamics at SIS energies is reasonably
described within the hadronic transport approach HSD when including
meson potentials \cite{CBPR98}, this no longer holds at AGS energies
\cite{CBPR98}.  The heaviest system studied here is Au~+~Au at $\approx$
11 A$\cdot$GeV.  The calculated $\pi^-$, $K^+$ and $K^-$ rapidity
spectra for central (0-10\%) reactions ($b \leq$ 2 fm) are displayed in
Fig.~\ref{Fig3} in comparison to the data from Ref. \cite{E866}. The
solid histograms correspond to the 'bare mass' scenario and
underestimate the data strongly whereas the dashed histograms are
obtained for $\alpha_K = -0.06$ and $\alpha_{\bar{K}}$ = 0.24,
respectively. Whereas the $K^-$ yield is almost reproduced in the
latter scheme, the $K^+$ yield is still underestimated as in case of
the Si~+~Au system at 14.6 A$\cdot$GeV \cite{Geiss}.

\begin{figure}[t]
\phantom{a}\vspace*{-10mm}
\psfig{figure=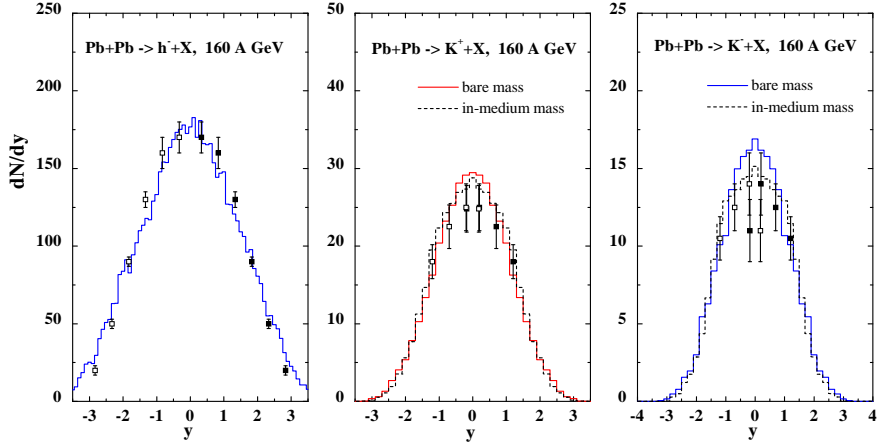,width=12.5cm}
\vspace*{-14mm}
\caption{The calculated $h^-$ (l.h.s.), $K^+$ (middle) and $K^-$ (r.h.s.)
rapidity spectra for central Pb + Pb reactions at 160 A$\cdot$GeV in
comparison to the experimental data. The solid histograms are
obtained without meson potentials whereas the dashed histograms represent
calculations with the same potentials as at SIS energies (see text).}
\label{Fig4}
\end{figure}

We now step on to SPS energies.  The calculated results for the
negative hadron ($h^-$) (l.h.s.), kaon (middle) and antikaon (r.h.s.)
rapidity distributions for central collisions of Pb~+~Pb at 160
A$\cdot$GeV are shown in Fig.~\ref{Fig4} in comparison to the data from
\cite{Bormann}. The solid histograms correspond to the 'bare mass'
scenario whereas the dashed histograms reflect the 'in-medium mass'
case with $\alpha_K = -0.06$ and $\alpha_{\bar{K}}$ = 0.24. As for
S~+~S at 200 A$\cdot$GeV \cite{Geiss} the $h^-$, $K^+$ and $K^-$
distributions are reproduced rather well showing even a tendency for an
excess of kaons and antikaons in the calculations rather than missing
strangeness. The distribution in rapidity becomes slightly broader for
in-medium kaon masses (dashed histograms in Fig.~\ref{Fig4}), but both
scenarios are compatible with the present data. Kaon and antikaon
self-energies thus are hard to extract from data at SPS energies due to
the low sensitivity of the spectra on in-medium potentials. We mention
that the strangeness enhancement at SPS energies in nucleus-nucleus
collisions does not qualify as a signal for an intermediate quark-gluon
plasma (QGP) phase since the spectra are fully compatible with a
hadronic reaction scenario. However, the strangeness production at much
lower energy appears more promising \cite{Geiss}.

\begin{figure}[t]
\phantom{a}\vspace*{-10mm}\hspace*{10mm}
\psfig{figure=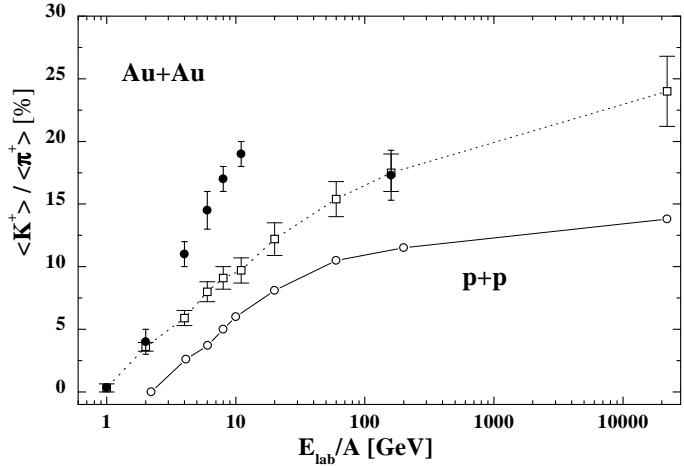,width=10cm}
\caption{The calculated $K^+/\pi^+$ ratio at midrapidity for central
Au + Au reactions (full squares) from SIS to RHIC energies in
comparison to the preliminary experimental data from 1 - 160 A$\cdot$GeV
and the corresponding ratio for p + p collisions (open circles) from the
HSD approach (see text). }
\label{Fig5}
\end{figure}

The E866 and E895 Collaborations recently  have measured Au~+~Au
collisions at 2,4,6 and 8 A$\cdot$GeV kinetic energy at the AGS
\cite{Ogilvi}.  Thus it is of particular interest to look for a {\em
discontinuity in the  excitation functions} for pion and kaon rapidity
distributions and to compare them to the hadronic HSD transport
approach.  In Fig. \ref{Fig5} the calculated $K^+/\pi^+$ ratios (open
squares) at midrapidity ($|y_{cm}| \leq$ 0.25) for central (b=2 fm)
Au~+~Au collisions at 1,2,4,6,8 and 11 A$\cdot$GeV and Pb + Pb
collisions at 160 A$\cdot$ are shown together with the preliminary data
(full dots). The ratio at midrapidity is slightly higher than the total
$K^+/\pi^+$ ratio, because the kaon rapidity distribution is narrower
than that of the pions.  While the scaled kaon yield at 1 and 2
A$\cdot$GeV (SIS energies) is well described in the HSD approach within
the errorbars, the experimental $K^+/\pi^+$ ratio at 4 A$\cdot$GeV is
underestimated already by a factor of 2 and increases up to roughly
19\% for 11 A$\cdot$GeV.  As mentioned before the calculated and
measured ratio coincide again at 160 A$\cdot$GeV.

The increase of the $K^+/\pi^+$ ratio with bombarding energy in the HSD
approach is much slower up to 11 A$\cdot$GeV in comparison to the data
and similar to the corresponding ratio for p + p collisions (open
circles) in HSD. The relative strangeness enhancement in the transport
approach for Au + Au compared to p + p is due to Fermi motion and
hadronic rescattering; obviously this hadronic rescattering scenario is
insufficient to describe the experimental excitation function.

\section{Summary}
Pions and $\eta$-mesons are found not to show sizeable in-medium
effects in their relative abundancy and spectra. The same holds for
kaons where only the data on kaon flow indicate a slightly repulsive
potential. Antikaons similar to antiprotons \cite{CBPR98,Sib98},
however, do show strong attractive potentials in the medium which is
seen most clearly at 'subthreshold' production energies in
nucleus-nucleus collisions. The magnitude of the $K^+, K^-$ potentials
seen at finite density is roughly in line with Lagrangian models based
on chiral perturbation theory \cite{Kaplan,waas} and relativistic
mean-field approaches \cite{Schaffner}.

An enhancement of the $K^+/\pi^+$ ratio in heavy-ion collisions
relative to p + p reactions is found due to hadronic rescattering both
with increasing system size and energy.  It should be emphasized that
this is expected within any hadronic model:  the average kinetic energy
and the particle density increases monotonically with incoming kinetic
energy of the  projectile while the life time of the fireball increases
with the system size. The excitation function in the $K^+/\pi^+$ ratio
from the hadronic transport approach has a similar slope in
nucleus-nucleus and p + p collisions (cf. Fig. 5) indicating a
monotonic increase of strangeness production with bombarding energy.
However, the experimental $K^+/\pi^+$ ratio for central Au + Au
collisions at midrapidity increases up to $\approx 19\%$ at 11
A$\cdot$GeV -- it is unknown if a local maximum will be reached at this
energy -- and decreases at SPS energies to $\approx 16.5\%$.  Such a
decrease of the scaled kaon yield from AGS to SPS energies is hard to
obtain in a hadronic transport model. On the contrary, the higher
temperatures and particle densities at SPS energies allways tend to
enhance the $K^+/\pi^+$ yield closer to its thermal equlibrium value of
$\approx 20-25\%$ \cite{BraunMu} at chemical freezout and temperatures
of $T\approx 150$ MeV.  Thus the steep rise of the strangeness yield
and its decrease indicates the presence of nonhadronic degrees of
freedom which might become important already at about 4 A$\cdot$GeV;
according to our present understanding such nonhadronic degrees of
freedom should be attributed to partons, i.e. quarks and gluons.

\end{document}